# Structural and Magnetic Properties of Sm(Co$_{0.7}$Fe$_{0.1}$Ni$_{0.12}$Zr$_{0.04}$B$_{0.04}$)$_{7.5}$ Melt Spun Isotropic and Anisotropic Ribbons

Sofoklis S. Makridis[1,*], Wei Tang[2]

[1] Department of Mechanical Engineering, University of Western Macedonia, GR-50100, Kozani, Greece

[2] Division of Materials Sciences & Engineering, Ames Lab of DOE, Iowa State University, Ames, IA 50011 USA

**Abstract.** We have investigated the structural and magnetic properties of Sm(Co$_{0.7}$Fe$_{0.1}$Ni$_{0.12}$Zr$_{0.04}$B$_{0.04}$)$_{7.5}$ melt spun ribbons. Samples were arc melted then melt spun at 37 m/s up to 55 m/s to obtain ribbon for powdering. Annealing has been performed in argon atmosphere for (30 to 75) min at (600 to 870) $^o$C. In as-spun ribbons the hexagonal SmCo$_7$ (TbCu$_7$-type of structure) of crystal structure has been determined from x-ray diffraction patterns, while fcc-Co has been identified as a secondary phase. After annealing, the 1:7 phase of the as-spun ribbons transforms into 2:17 and 1:5 phases. X-ray patterns for as-milled powders exhibit very broad peaks making it difficult to identify a precise structure but represent the 1:7 structure after annealing at low temperature (650 $^o$C). TEM analysis shows a homogeneous nanocrystalline microstructure with average grain size of (30 to 80) nm. Coercivity values of (15 to 27) kOe are obtained from hysteresis loops traced up to a field of 5 T. The coercivity decreases as temperature increases, but it maintains values higher than 5 kOe at 380 $^o$C. The maximum energy product at room temperature increases, as high as 7.2 MGOe, for melt-spun isotropic ribbons produced at higher wheel speeds. Anisotropic ribbons have a maximum energy product close to 12 MGOe.

## Introduction

Permanent magnets based on the Sm(Co,Fe,Cu,Zr)$_z$ alloys have high energy product and high Curie temperature. Optimization studies for use at high temperatures performed recently resulted in compositions with intrinsic coercivity (H$_c$) of 10 kOe at 450 $^o$C [1]. On these materials coercivity is obtained by a precipitation hardening process, which is comprised of a solid solution treatment at (1100 to 1200) $^o$C for (4 to 24) h, isothermal aging at 850 $^o$C for (10 to 24) h and slow cooling to 400 $^o$C at cooling rate <1 $^o$C/min. This prolonged and complicated heat treatment is necessary for the development of the typical cellular-lamellar microstructure, with rhombohedral Th$_2$Zn$_{17}$ (2:17R) type structure cells, Cu-rich cell boundaries and Zr-rich lamellae perpendicular to the c axis. The presence of Cu and Zr is considered a requirement for the formation of the complex microstructure. Substitution of Ni for Cu does not prevent the formation of the cellular - lamellar microstructure, but the resulting coercivity is very low at room temperature, which increases with temperature [2]. In melt-spun Sm(Co,Fe,Cu,Zr)$_z$ alloys the cellular-lamellar microstructure and high coercivity (28 kOe) has been achieved simply by slow cooling the ribbons from (850 to 400) °C, without the conventional solid solution and aging treatments [3]. Precipitation hardening with cellular features and H$_c$ up to 10 kOe has also been obtained in melt-spun ribbons after a short annealing time [4].

Alternative processing routes for nanostructured magnets with high coercivity were explored in several recent studies. The 1:7 or 2:17 stoichiometries have been prepared by mechanical milling and subsequent annealing yielding samples with with coercivities up to 20 kOe to 25 kOe [5-7].

---

[1] Corresponding author: Sofoklis Makridis, email: smakridis@uowm.gr

Rapid solidification by melt spinning and short heat treatment has produced lower $H_c$ values, from 5 to 14 kOe [8-11, 4]. Much higher coercive fields (16 to 27) kOe have been obtained, in boron substituted $Sm(Co_{bal}Fe_{0.1}Cu_{0.12}Zr_{0.04}B_x)_{7.5}$ as-spun ribbons or after short annealing [12,13]. These alloys are typical of bulk precipitation hardened Sm-Co magnets, but the addition of B and the different synthesis process leads to a different microstructure. Many other groups have focused on this direction [14-28]. In our series, the samples with boron content of 4 at. %, melt-spun at moderate wheel speed (~40 m/s), showed very good magnetic properties and provided a promising starting point for further optimization studies. The effect of Ni in place of the non-magnetic Cu has already been published in ref. [29]. In the present work, the effect of the wheel speed, the annealing conditions, and the texture on the structural and magnetic properties of $Sm(Co_{0.7}Fe_{0.1}Ni_{0.12}Zr_{0.04}B_{0.04})_{7.5}$ ribbons were investigated in order to optimize their performance for room or high temperature applications.

**Experimental procedure**

Master alloys with $Sm(Co_{0.7}Fe_{0.1}Ni_{0.12}Zr_{0.04}B_{0.04})_{7.5}$ composition were prepared by arc-melting in an Ar atmosphere. To compensate for the Sm losses during processing, an excess of (~5 to 10) wt. % Sm was added. Ribbons were obtained from the master alloys by melt spinning at wheel speeds from 37.5 to 55 m/s using a quartz tube with an orifice diameter of 0.5 mm under 2 atm pressure of 99.999 % pure argon. The wheel velocity was determined by using a stroboscope. The ribbons were wrapped with tantalum foil and sealed in a quartz tube after three vacuum purges with pure argon to avoid oxidization, and then annealed in the temperature range of 600 °C to 870 °C for 30 to 75 min. In addition, a RETSCH PM400 high energy planetary ball mill was used to mill the ribbons to produce fine powder. The sample of powdered ribbons was mixed with paste and the mixture was aligned under a magnetic field of 2 T in order to make anisotropic magnetically oriented sample. The phases in as-spun and annealed ribbons were determined by x-ray powder diffraction (XRD) using Fe-$K_\alpha$ radiation. The magnetic measurements were performed with a SQUID magnetometer with a maximum applied field of 50 kOe. The high temperature measurements were performed in a vibrating sample magnetometer (VSM) with a maximum field of 20 kOe from room temperature up to 600 °C.

**Results and discussion**

X-ray diffraction patterns of as spun ribbons with different wheel speeds are shown in Fig. 1. The patterns for all as spun samples are identified as a 1:7 structure and refined as $TbCu_7$-type of structure. The 1:7 is a metastable phase, commonly considered as a disordered 2:17 rhombohedral phase, which is dominant to develop final microstructure and magnetic properties for the ribbons. The lattice parameters for the 1:7 ($SmCo_7$) phase are determined as a=b≈4.95 Å, c≈4.07 Å after Rietveld analysis. The diffraction peaks (111), (200) and (220) of fcc-Co for $\lambda_{(Fe-K\alpha)}$ correspond at angles ~56°, ~66° and ~101°, respectively and they can vaguely be seen in some as-spun diffraction patterns [30].

After the as-spun ribbons are annealed at 810 °C for 1 h and subsequently quenched, a phase transformation of the $TbCu_7$-type of structure (1:7) into 2:17 rhombohedral and 1:5 structures occurs, as shown in patterns of Fig. 2. The rhombohedral 2:17 type structure can be derived from the $CaCu_5$ (1:5) type structure if a dumbbell pair of transition metal atoms substitutes for 1/3 of rare earth atoms. If the distribution of dumbbell pairs is random, then the structure is not the 2:17 type, but the 1:7. Annealing induces an ordered distribution of dumbbell pairs and the formation of 2:17R and 1:5 structures. Traces of fcc-Co are also found in short-time annealed samples.

In order to modify the microstructure of as spun ribbons, while avoiding the phase transformation by keeping one high anisotropic phase (1:7), some of as spun ribbons samples were annealed below 750 °C. It was found that the annealing around 600 °C for 30 min / 75 min leads to the formation of a more homogeneous microstructure. The properties of the ribbons spun at 39 m/s and annealed at 810 °C and 600 °C were thoroughly investigated in our previous paper [29]. It has been shown in that work that low temperature annealing is more effective on the hard magnetic properties.

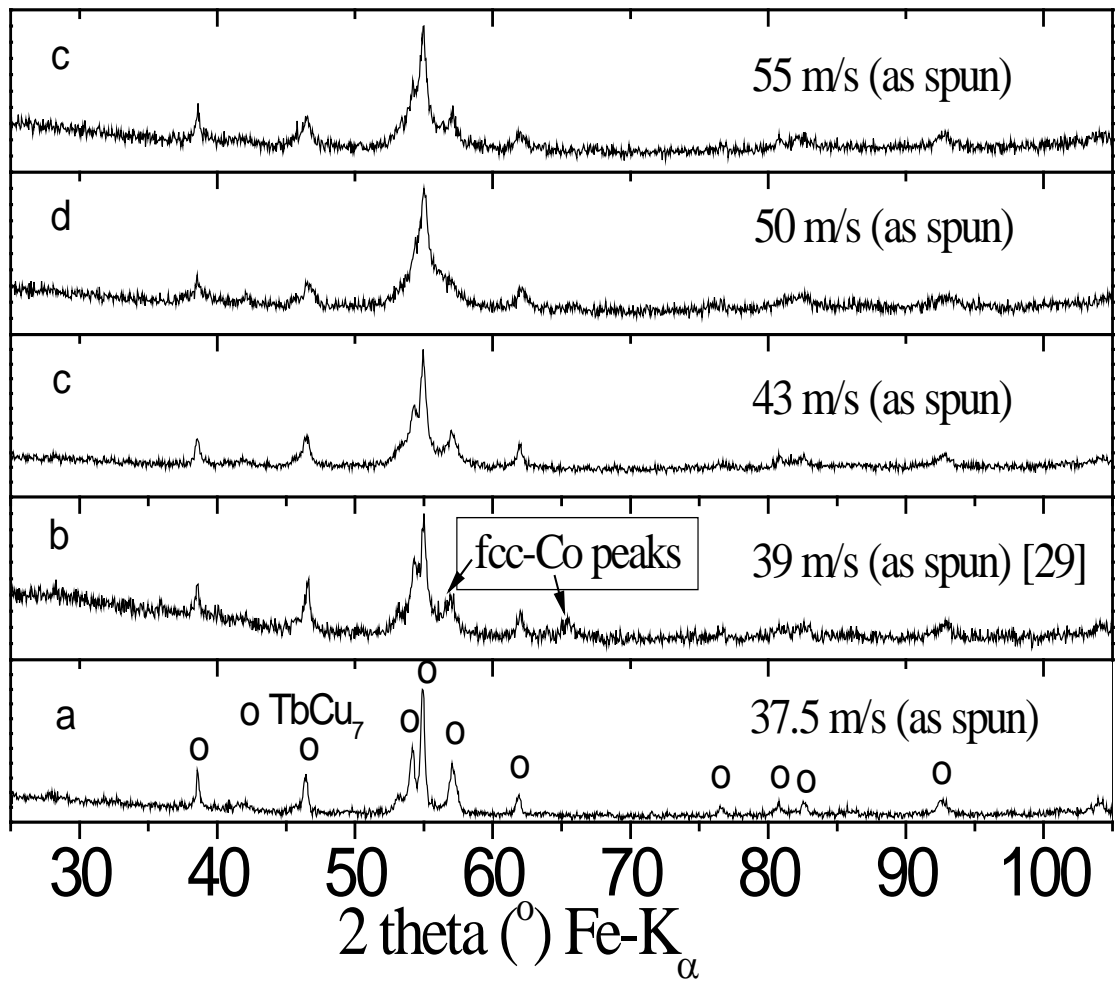

**Fig. 1.** X-ray diffraction patterns of as spun ribbons prepared at different wheel speeds (a)-(e). The crystal structure of the as spun ribbons is TbCu$_7$ type of structure (symbol: o)

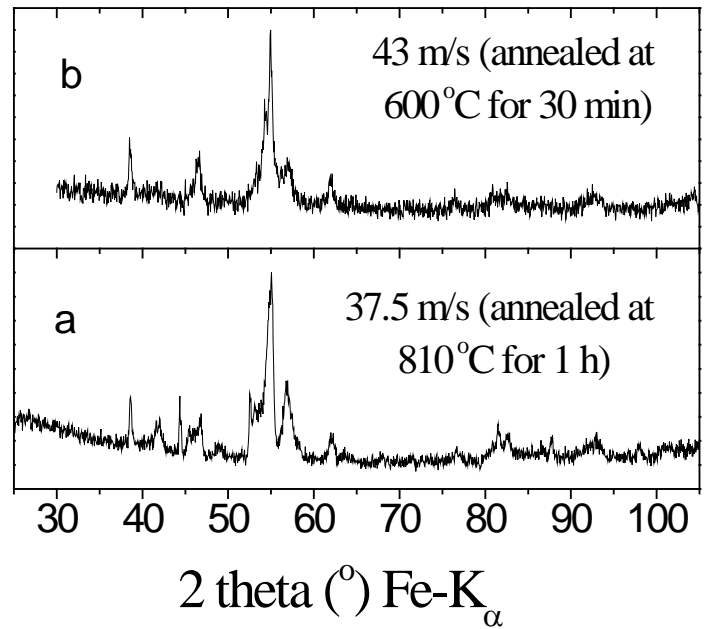

**Fig. 2.** X-ray diffraction patterns of annealed ribbons at different temperatures (a)-(b)

Fig. 3 shows the room temperature initial magnetization and demagnetization curves M(H) of as-spun ribbons produced at 37.5 m/s to 50 m/s. The maximum applied magnetic field was 50 kOe. All as-spun ribbons have coercive fields equal to or larger than 15 kOe. The highest coercivity was obtained in the ribbon sample melt spun at 43 m/s. As spun ribbons prepared at 55 m/s were also measured at a maximum applied field of 20 kOe. The hysteresis loop of the sample exhibits (not shown in Fig. 3) a coercivity of 12.2 kOe with a steep step on the demagnetization curve. All of as-spun samples primarily consist of the magnetically hard phase. There exists a small amount of soft phases, like fcc-Co (as identified from the XRD patterns), which can result in magnetically decoupling the hard phase or in microstructural non-uniformities, and thus result in the formation of the step ("knee") on the demagnetization curves at the reverse fields close to zero. In comparison, the sample melt spun at 39 m/s obtains better reduced remanence ($m_r = M_r/M_s$) and coercive field ($H_c$) values, as well as a more ideal shape of the demagnetization curve, as shown in Fig. 3b.

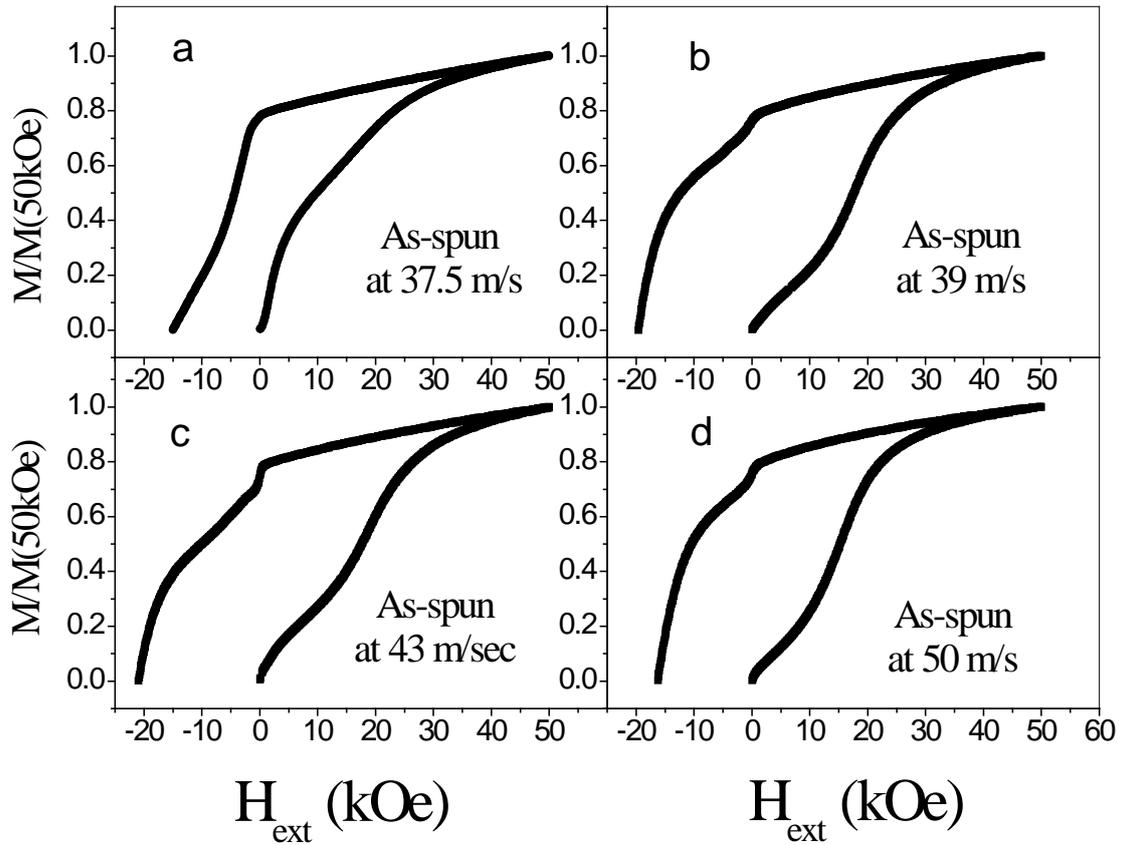

**Fig. 3.** Magnetic hysteresis loops for melt spun ribbons at (a). 37.5 m/s, (b). 39 m/s, (c).43 m/s and d). 50 m/s, respectively.

It is observed that the initial curves of the samples obtained by melt spinning at 39, 43, and 50 m/s (except for 37.5 m/s), as shown in Fig. 3b, 3c, and 3d, have a particular shape which is often observed in magnets with strong exchange-coupling between nanograins and with the "pinning" type coercivity mechanism. The initial curve for the sample produced at 37.5 m/s exhibits a larger slope (see Fig. 3a), suggesting the coercivity nucleation mechanism in the sample. The saturation magnetizations of as-spun ribbons prepared at 37.5, 39, 43, and 50 m/s are measured as 57, 62, 69, and 76 emu/g, respectively. It is seen that the saturation magnetization increases with increasing wheel speed. The as-spun ribbons produced at different wheel speeds were annealed at different temperature and time conditions. The hysteresis loops for the annealed ribbons are shown in Fig. 4. The maximum energy product $(BH)_{max}$ values were calculated from these original loops without the correction of demagnetization factors which means that the real $(BH)_{max}$ values should be higher than the calculated ones. For the sample melt spun at 43 m/s and annealed at 600 °C for 30 min, the $H_c$ is 21 kOe, $m_r$=0.76 and $(BH)_{max}$=7.2 MGOe, but the hysteresis curve is not an ideal shape (Fig. 4c). After the sample melt spun at 39 m/s and annealed at 600°C for 30 min is powdered and aligned, the coercivity is reduced to 16.7 kOe, while $m_r$ and $(BH)_{max}$ are increased to 0.9 and 11.9 MGOe, respectively, as shown in Fig. 4d.

The thermal stability of the coercive field is very important for high temperature applications. The high temperature coercive fields for as-spun and annealed samples were measured with a maximum applied field of 20 kOe in the temperature range of 70 to 550 °C. Fig. 5 shows the temperature dependence of coercivity for selected ribbons. The coercivity decreases with increasing temperature for all samples. Compared to the samples obtained by different melt spinning velocities and annealing temperatures, it was found that the samples can achieve better magnetic properties by melt spinning at velocities higher than 39 m/s, and annealing at lower temperatures. A highest coercivity of 5 kOe at 380 °C was obtained in the sample melt spun at 43 m/s and annealed at 600 °C for 30 min (see Fig. 3e). In addition, the coercivity change with temperature is reversible, as shown in Fig. 3e, 3f, the coercivity is nearly the same after heating up to 450 °C and cooling back to 100 °C [29]. The sample shown in Fig 3e or 3f has the best thermal stability in the studies. The Table I has detailed information about the magnetic and structural properties of the investigated samples.

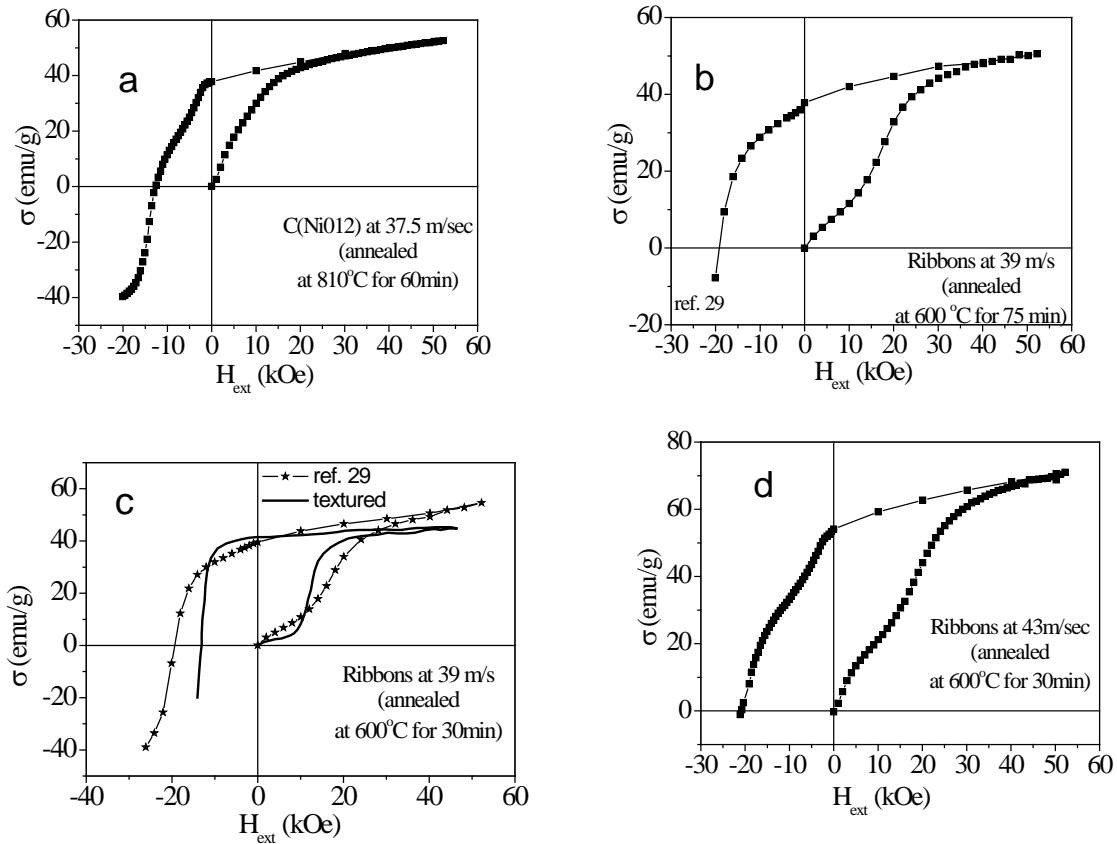

**Fig. 4.** Magnetic hysteresis loops for annealed ribbons at (a).37.5, (b).39 – 75 min [29], (c).39 – 30 min [29] and the hysteresis loop of the textured sample at 39 m/s, and (d). 43 m/s.

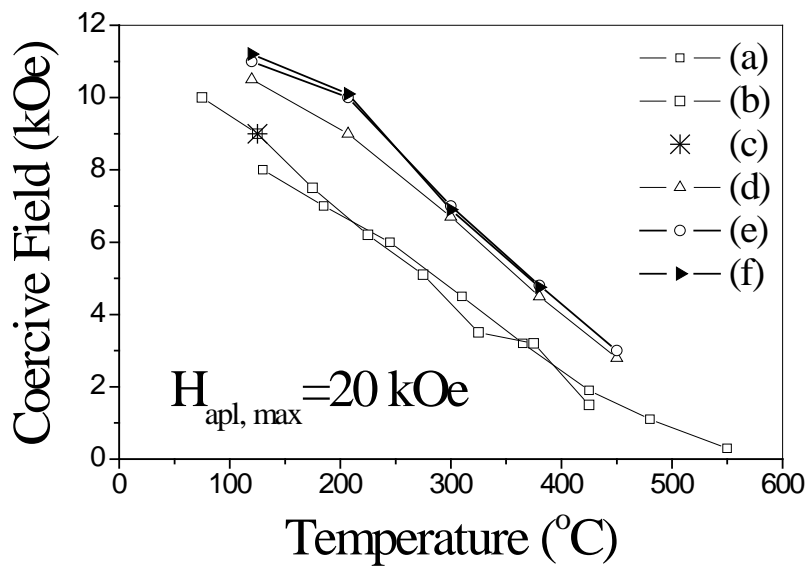

**Fig. 5.** Coercivity as a function of temperature for melt spun ribbons: (a) as spun at 39 m/s [ref. 29], (b) spun at 39 m/s and annealed at 810 °C for 1 h [ref. 29], (c) spun at 39 m/s and annealed at 810 °C for 1 h after cooling [ref. 29], (d) as spun at 37.5 m/s, (e) heating: spun at 43 m/s and annealed at 600 °C for 30 min, (f) cooling: spun at 43 m/s and annealed at 600 °C for 30 min.

Table I. Summarizing the properties of investigated materials based in $Sm(Co_{0.7}Fe_{0.1}Ni_{0.12}Zr_{0.04}B_{0.04})_{7.5}$ stoichiometry

| Sample's speed (m/s) | Structure (with fcc-Co) | $H_c$ (kOe) | $m_r$ | $M_s$ or $(BH)_{max}$ |
|---|---|---|---|---|
| 37.5 | $TbCu_7$ type | 16 | 0.76 | 57 emu/g |
| 37.5 annealed 810 °C | $Th_2Zn_{17}/CaCu_5$ type | 14 | 0.7 | 53.2 emu/g |
| 39 | $TbCu_7$ type | 19.5 | 0.78 | 62 emu/g |
| 39 annealed 600 °C (as-spun) | $TbCu_7$ type | 21 | 0.76 | 53 emu/g |
| 39 annealed 600 °C (textured) | $TbCu_7$ type | 16.7 | 0.9 | 11.9 MGOe |
| 43 | $TbCu_7$ type | 21 | 0.77 | 69 |
| 43 annealed 750 °C | $TbCu_7$ type | 21 | 0.76 | 7.6 MGOe |
| 50 | $TbCu_7$ type | 16 | 0.75 | 76 |
| 55 | $TbCu_7$ type | 12.2 | 0.76 | 52 |

TEM image of the best sample spun at 43 m/s and annealed at 600 °C for 30 min was studied and shown in Fig. 6. The TEM image exhibits a homogeneous nanocrystalline microstructure with average grain size of 30-80 nm. This should be responsible for the good exchange coupling between the hard and soft phases. The microstructural analysis shows that most of the grains are small but some bigger grains are also present. It is well known that large grains or agglomerates can directly lead to weaker exchange coupling between the hard and the soft fcc-Co grains. The weak interactions are responsible for the presence of the small step close to zero applied field in the demagnetization curves, as shown in Fig. 3 and Fig. 4.

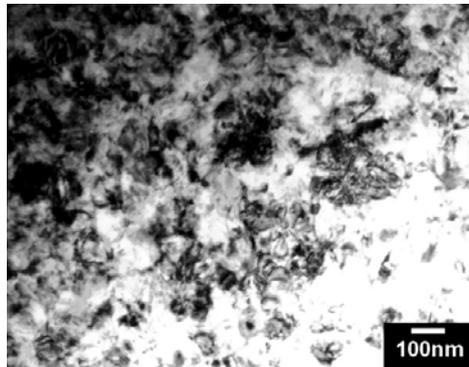

**Fig. 6.** TEM image of the ribbons spun at 43 m/s annealed at 600 °C for 30 min

## Summary

The melt spun ribbons with $Sm(Co_{0.7}Fe_{0.1}Ni_{0.12}Zr_{0.04}B_{0.04})_{7.5}$ composition were systematically examined. The structural characteristics of the 1:7 phase are important for efficiency of the magnetic material. Hard-soft exchange coupling results in high coercivity and good loop squareness. Wheel speed and annealing temperatures are crucial parameters for developing higher coercivity and better temperature performance. The reduced magnetization and loop squareness were further improved by aligning powder ribbons. The highest maximum energy product values were obtained in the aligned sample. The coercivity decreases with increasing temperature and is reversible at the range of room temperature to 450 °C. The best sample retains a coercivity of 5 kOe at 380 °C. Moreover, the discussed maximum energy product in Ref [29] - and Ref [30] therein - which is associated to this research work was increased notable, more than 70 %, though the anisotropic treatment.

**Acknowledgements**

Dr. G. Litsardakis provided access to the SIEFERT X-ray diffractometer. HITEMAG European project and DARPA Metamaterials program are gratefully acknowledged.

**Graphical Abstract**

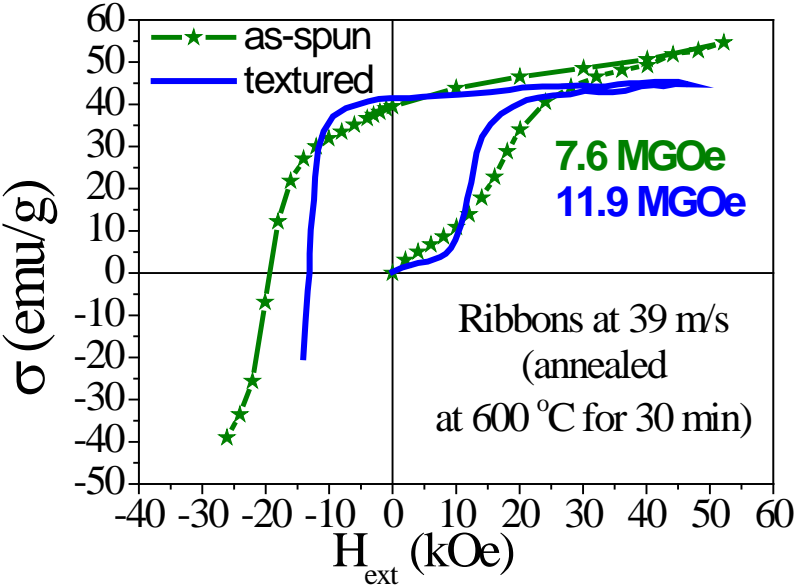